\documentclass[amsmath, amssymb, superscriptaddress]{revtex4-2}
\usepackage{graphicx}

\graphicspath{{.}}

\newcommand{\BP}[1][ ]{Bloch{#1}point}

\usepackage[
pdftitle={Controlling stable Bloch points with electric currents},
pdfauthor={Martin Lang, Swapneel Amit Pathak, Samuel J. R. Holt, Marijan Beg, and Hans Fangohr},
colorlinks=True,linkcolor=black,citecolor=black
]{hyperref}

\usepackage{orcidlink}
\begin{document}

\title{Controlling stable Bloch points with electric currents}
\author{Martin Lang\,\orcidlink{0000-0001-7104-7867}}
\email{martin.lang@mpsd.mpg.de}
\affiliation{Faculty of Engineering and Physical Sciences, University of
  Southampton, Southampton, SO17 1BJ, United Kingdom}
\affiliation{Max Planck Institute for the Structure and Dynamics of Matter,
  Luruper Chaussee 149, 22761 Hamburg, Germany}

\author{Swapneel Amit Pathak\,\orcidlink{0000-0003-3840-955X}}
\affiliation{Max Planck Institute for the Structure and Dynamics of Matter,
  Luruper Chaussee 149, 22761 Hamburg, Germany}

\author{Samuel J. R. Holt\,\orcidlink{0000-0003-3323-8958}}
\affiliation{Max Planck Institute for the Structure and Dynamics of Matter,
  Luruper Chaussee 149, 22761 Hamburg, Germany}

\author{Marijan Beg\,\orcidlink{0000-0002-6670-3994}}
\affiliation{Department of Earth Science and Engineering, Imperial College
  London, London SW7 2AZ, United Kingdom}
\affiliation{Faculty of Engineering and Physical Sciences, University of
  Southampton, Southampton, SO17 1BJ, United Kingdom}

\author{Hans Fangohr\,\orcidlink{0000-0001-5494-7193}}
\affiliation{Max Planck Institute for the Structure and Dynamics of Matter,
  Luruper Chaussee 149, 22761 Hamburg, Germany}
\affiliation{Faculty of Engineering and Physical Sciences, University of
  Southampton, Southampton, SO17 1BJ, United Kingdom}
\affiliation{Center for Free-Electron Laser Science, Luruper Chaussee 149, 22761
  Hamburg, Germany}

\begin{abstract}
  The \BP{} is a point singularity in the magnetisation configuration, where the
  magnetisation vanishes. It can exist as an equilibrium configuration and plays
  an important role in many magnetisation reversal processes. In the present work, we
  focus on manipulating \BP{}s in a system that can host stable \BP{}s---a
  two-layer FeGe nanostrip with opposite chirality of the two layers. We drive
  \BP{}s using spin-transfer torques and find that \BP{}s can move collectively
  without any Hall effect and report that \BP{}s are repelled from the sample boundaries and
  each other. We study pinning of \BP{}s at wedge-shaped constrictions (notches)
  in the nanostrip and demonstrate that arrays of \BP{}s can be moved past a
  series of notches in a controlled manner by applying consecutive current
  pulses of different strength. Finally, we simulate a T-shaped geometry and
  demonstrate that a \BP{} can be moved along different paths by applying
  current between suitable strip ends.
\end{abstract}

\maketitle

Bloch points~\cite{Feldtkeller1965, Doring1968} are point singularities that can
be observed in magnetic systems. They have first been studied in the context of
bubble memories~\cite{Malozemoff1979}. Bloch points play an important role in
many dynamical processes, such as vortex-antivortex
annihilation~\cite{Hertel2006}, vortex core reversal~\cite{Thiaville2003}, and
skyrmion reversal~\cite{Beg2015}. Recently, their static microscopic
structure~\cite{Donnelly2017} and their dynamics~\cite{Im2019} were measured
experimentally.

Beg \emph{et al.}~\cite{Beg2019} demonstrated that a stable \BP{} can exist in a
thin helimagnetic cylinder consisting of two layers with opposite chirality. The
\BP{} in this system is of circulating type and exists in two different
configurations, head-to-head (HH) and tail-to-tail (TT). In a recent
work~\cite{Lang2023}, we have demonstrated that rectangular two-layer nanostrips
with opposite chirality can host multiple \BP{}s, and in a nanostrip of suitable
size any combination of \BP{}s of the two types can be in equilibrium, which is
necessary for encoding data with \BP{}s.

In this work, we study the motion of one or multiple \BP{}s under applied
spin-transfer torques in a system consisting of two layers with opposite
chirality. We find that the \BP{} moves without any Hall effect in the two-layer
geometry. Multiple \BP{}s move collectively, independent of their type or
arrangement. We study the effect of geometry variations of the nanostrip by
removing the magnetic material in one or multiple notches at the edge of the
nanostrip. Finally, we simulate the motion of a \BP{} in a T-shaped geometry and
show that we can control the motion of the \BP{} with the applied current.
Depending on the current direction, the \BP{} can move along different paths,
i.e.\ between arbitrary ends of the structure.

\section{Results}
\label{sec:results}
\subsection{Rectangular strips}

\begin{figure}
  \includegraphics[width=\linewidth]{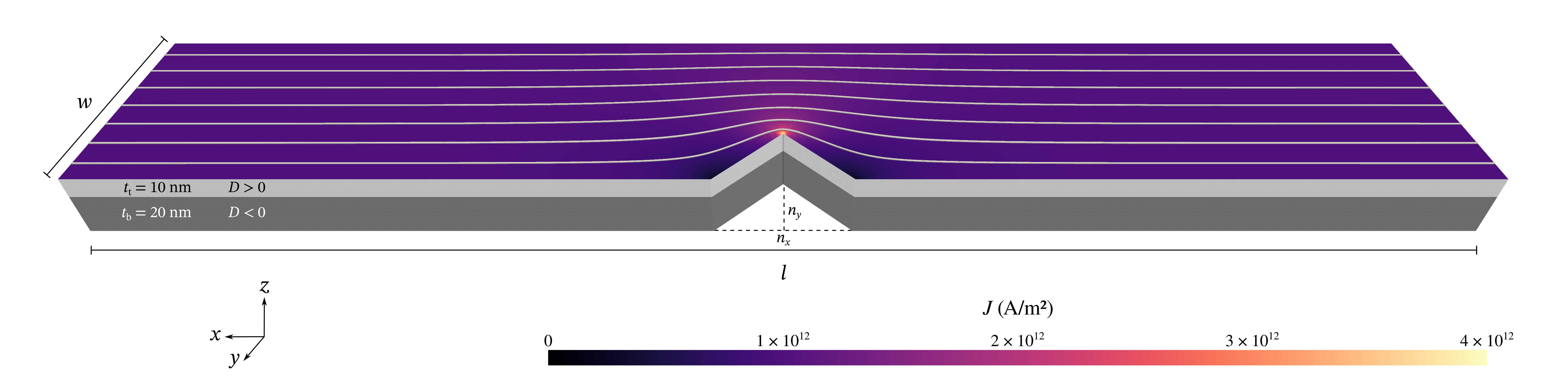}
  \caption{\label{fig:strip-geometry}Geometry of the nanostrips studied in this
    work. All nanostrips consist of two layers with opposite chirality (light
    and dark grey), length $l$ and width $w$ are adjusted as required. In a
    later part of this work we also study the effect of notches with width~$n_x$
    and depth~$n_y$ in the strip. They extend throughout the whole sample
    thickness, as shown here. The colour and the streamlines on the top surface
    show the current density and direction in the nanostrip with a notch.}
\end{figure}

First, we focus on driving \BP{}s in rectangular strips. We simulate nanostrips
with length $l=1500\,\mathrm{nm}$ or $l=2000\,\mathrm{nm}$ and width
$w=100\,\mathrm{nm}$. The nanostrips consist of two layers with opposite material
chirality, i.e. opposite sign of the Dzyaloshinkii-Moriya constant. The bottom
layer has thickness $t_{\mathrm{b}}=20\,\mathrm{nm}$, the top layer thickness
$t_{\mathrm{t}}=10\,\mathrm{nm}$. We show the geometry in
Fig.~\ref{fig:strip-geometry}, the two layers are indicated with dark and light
grey. The figure shows a nanostrip with an additional notch studied later in
this work. The top surface shows the corresponding simulated current density.

\begin{figure}
\includegraphics[width=\linewidth]{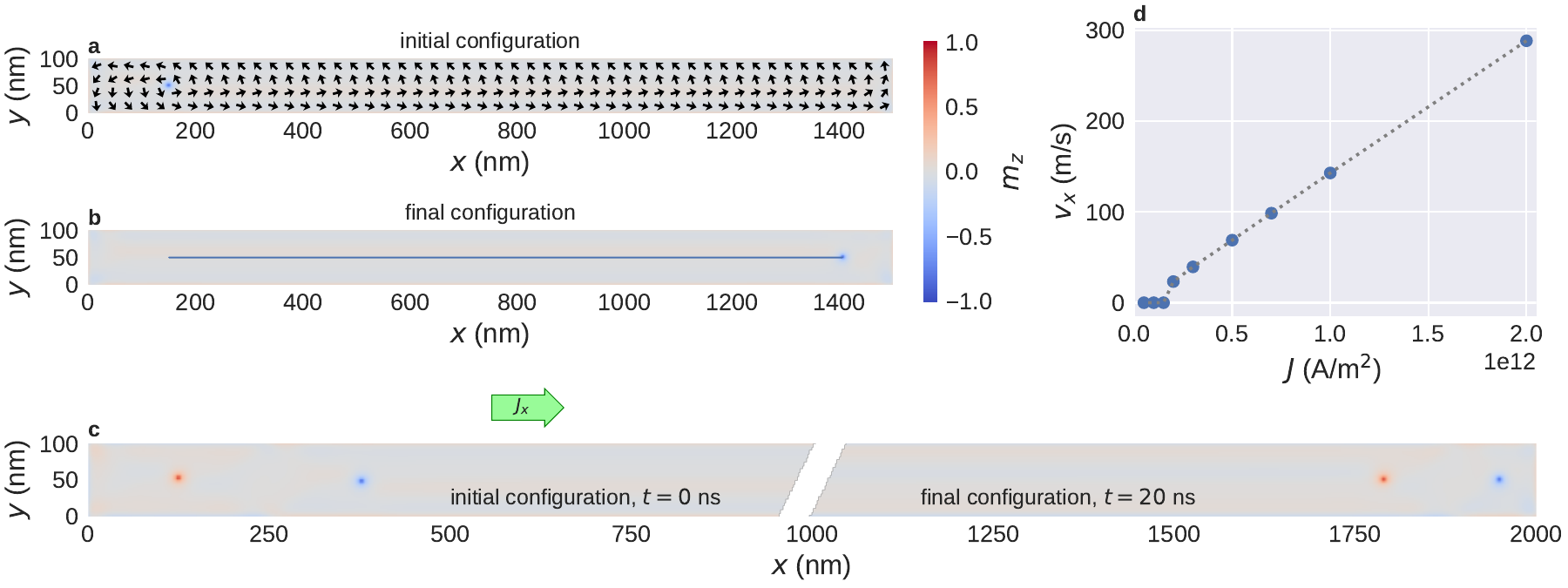}
\caption{\label{fig:driving-uniform}Movement of \BP{}s with applied spin current
  through uniform nanostrips. (a, b) Movement of a single \BP{} with current
  density $J=0.5\times 10^{12}\,\mathrm{A/m}^{2}$. The \BP{} is initialised near
  the left sample edge~(a). When applying current, the \BP{} moves in $+x$
  direction until it reaches the right sample edge~(b). (c)~Movement of two
  \BP{}s of opposite type. (d)~Speed of a single \BP{} depending on the current
  strength. A linear speed increase can be observed above a certain depinning
  threshold.}
\end{figure}

We begin by initialising the system and relaxing it to a state that contains one
or two \BP{}s near the left sample boundary and apply current in $+x$~direction.
Figure~\ref{fig:driving-uniform}a shows an initial configuration with a single
\BP{} located at $x\approx 150\,\mathrm{nm}$ in a $1500\,\mathrm{nm}$ strip. The
cross-section shows the magnetisation above the interface of the two layers,
colour indicates the out-of-plane component~$m_z$, and arrows the in-plane
components~$m_x$ and~$m_y$. When current is applied in $+x$~direction, the \BP{}
moves along the current direction until it reaches the right sample boundary,
where it stops due to the repulsion from the sample edge.
Figure~\ref{fig:driving-uniform}b shows the final configuration with the \BP{}
at $x\approx1400\,\mathrm{nm}$. The blue line shows the path of the \BP{}. We
can see that the \BP{} moves in a straight line and is not subject to any Hall
effect, independent of the strength of the applied current. This is
qualitatively different from the \BP{} in a chiral bobber studied in
Ref.~\onlinecite{Gong2021}, where the whole object is subject to a
current-dependent Hall effect. We have also performed simulations for much wider
strips ($w=600\,\mathrm{nm}$) to verify that the straight motion is not caused
by the repulsion from the sample boundaries in $y$~direction.

Figure~\ref{fig:driving-uniform}c shows an example for two \BP{}s in the
configuration tail-to-tail and head-to-head in a $2000\,\mathrm{nm}$ strip. The
left part shows the initial configuration with
$x_{\mathrm{TT,i}}\approx125\,\mathrm{nm}$ and
$x_{\mathrm{HH,i}}\approx375\,\mathrm{nm}$. The right part shows the final
configuration with $x_{\mathrm{TT,f}}\approx1800\,\mathrm{nm}$ and
$x_{\mathrm{HH,f}}\approx1950\,\mathrm{nm}$. Both \BP{}s move along the current
direction in a straight line. The \BP[-]{} type does not affect the motion.
We obtain qualitatively similar results for configurations containing two \BP{}s
of the same type and for configurations containing more than two \BP{}s.

Figure~\ref{fig:driving-uniform}d shows the average speed of a single \BP{}
as a function of the applied current density. We can observe pinning for small
current densities and linearly increasing velocity for larger current densities.
We cannot estimate the exact depinning current because it depends on the
discretisation.
Similar pinning behaviour of the \BP{} has been observed in other
micromagnetic~\cite{Thiaville2003, Im2019, Gong2021} and
atomistic~\cite{Kim2013, Hertel2015} simulations before. In the remainder of
this work we focus on larger current densities above the depinning threshold.

\subsection{Nanostrip with one notch}

\begin{figure}
  \includegraphics[width=\linewidth]{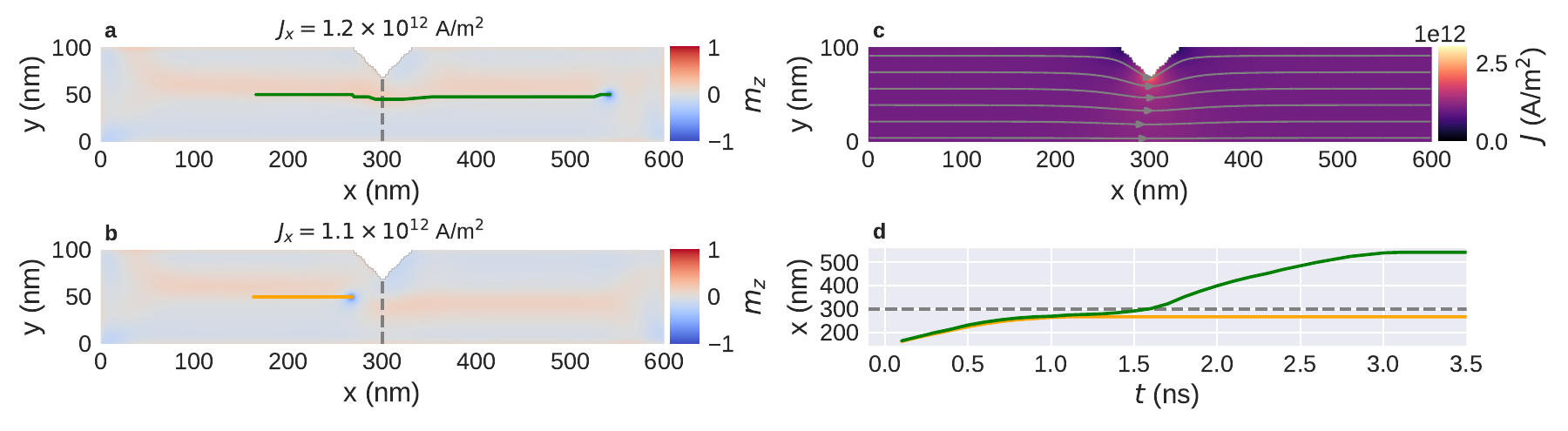}
  \caption{\label{fig:single-notch-configurations}Motion of a \BP{} in a
    nanostrip ($l=600\,\mathrm{nm}$, $w=100\,\mathrm{nm}$) with a single notch
    under applied current. (a, b)~Final configuration and \BP[-]{} trajectory for two
    different current strengths: (a)~the \BP{} can move past the notch for
    $J_x=1.2\times10^{12}\,\mathrm{A/m}{}^2$; (b)~the \BP{} is trapped at the
    notch for $J_x=1.1\times10^{12}\,\mathrm{A/m}{}^2$. (c)~Simulated current
    profile. (d)~Position of the \BP{}s in (a) and (b) as a function of
    simulation time.}
\end{figure}

Next, we study the motion of a single \BP{} in a nanostrip with a notch. We
simulate a strip with $l=600\,\mathrm{nm}$ with a wedge-shaped notch at
$x=300\,\mathrm{nm}$, extending through the sample in $z$~direction
(Fig.~\ref{fig:strip-geometry}). The notch tip is located at $y=70\,\mathrm{nm}$
($n_y=30\,\mathrm{nm}$) and the opening angle is 90° ($n_x=60\,\mathrm{nm}$). An $xy$~cross-section is
shown in Fig.~\ref{fig:single-notch-configurations}. We use a finite-element
simulation to compute the non-uniform current density profile in this geometry,
more details are provided in the Methods section. The resulting current profile
is shown in Fig.~\ref{fig:single-notch-configurations}c. Near the notch we can
observe a variation in the current density with the maximum at the tip of the
notch.

The simulated current densities on the order of $10^{12}\,\mathrm{A/m}^{2}$
would lead to a significant temperature increase due to Joule heating. A
detailed study~\cite{Fangohr2011} shows that the material, pulse duration, and
cooling from the substrate play important roles in the control of the
temperature. In this prototype study we ignore all temperature-related
effects and possible engineering efforts which would need to be addressed for
higher technical readiness levels.

First, we focus on two current densities
$J_{x,\mathrm{a}}=1.2\times10^{12}\,\mathrm{A/m}{}^2$ and
$J_{x,\mathrm{b}}=1.1\times10^{12}\,\mathrm{A/m}{}^2$. We initialise the systems
with a \BP{} at $x_{\mathrm{i}}\approx165\,\mathrm{nm}$. The final
configurations with applied current are shown in
Fig.~\ref{fig:single-notch-configurations}a and~b for $J_{x,\mathrm{a}}$ and
$J_{x,\mathrm{b}}$, respectively. The solid lines show the \BP[-]{} path from
its initial to its final configuration.
Figure~\ref{fig:single-notch-configurations}d shows the $x$~position of the
\BP{} in the two simulations as a function of simulation time.

The current density $J_{x,\mathrm{b}}$
(Fig.~\ref{fig:single-notch-configurations}b) is not strong enough to push the
\BP{} past the notch. Instead, the \BP{} gets stuck near the notch at a final
position $x_{\mathrm{f,b}}\approx270\,\mathrm{nm}$, to which it moves in a
straight line without any deflection in $y$~direction. As a function of time
(Fig.~\ref{fig:single-notch-configurations}d), we can see a slow-down as the
\BP{} approaches the notch. The motion stops around $t=1\,\mathrm{ns}$: the
applied current cannot push the \BP{} further against the restoring force from
the notch constriction.

The current density $J_{x,\mathrm{a}}$
(Fig.~\ref{fig:single-notch-configurations}a) is strong enough to push the \BP{}
past the notch, and the \BP{} stops at $x_{\mathrm{f,a}}\approx550\,\mathrm{nm}$
at around $t=3\,\mathrm{ns}$ due to the edge repulsion of the sample at
$x=600\,\mathrm{nm}$. Near the notch, we can see a small displacement in the
$-y$~direction, away from the tip of the notch. In the time-resolved data
(Fig.~\ref{fig:single-notch-configurations}d), we can see a slow-down of the
\BP{} in front of the notch, very similar to the results for $J_{x,\mathrm{a}}$
up to $t\approx1\,\mathrm{ns}$. The small deviations of the two curves for \BP[-]{}
positions before the notch are caused by the slightly different current
strengths. However, for the stronger current, the \BP{} continues to move
towards the notch for $t>1\,\mathrm{ns}$. We can see a slight speed increase
around $t=1.3\,\mathrm{ns}$, and the \BP{} passes the tip of the notch at around
$t=1.6\,\mathrm{ns}$ (and $x=300\,\mathrm{nm}$ shown as a dashed line in
Fig.~\ref{fig:single-notch-configurations}). After passing the notch, we can see
a strong speed increase and a fast motion in $+x$~direction until the \BP{}
approaches the right sample boundary, where it slows down and eventually stops.

\begin{figure}
  \includegraphics[width=\linewidth]{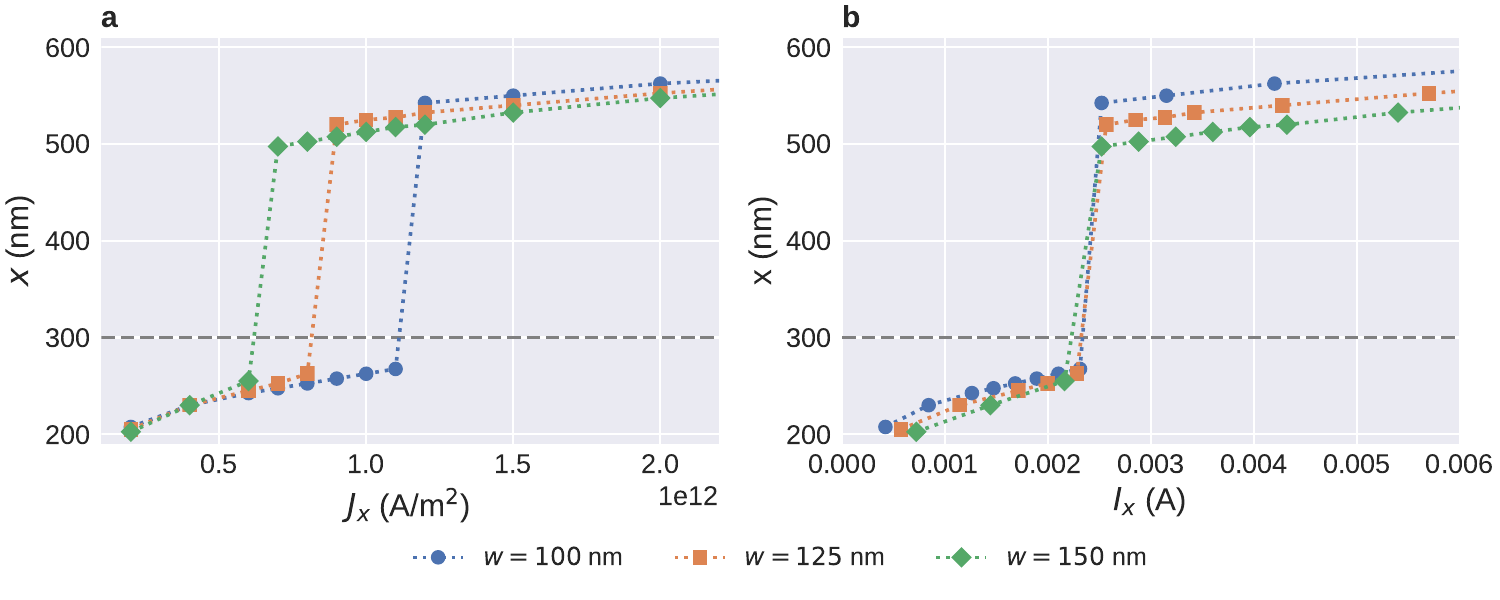}
  \caption{\label{fig:single-notch-current-density}Final equilibrium position of
  a \BP{} in a strip with a notch as a function of applied current for three
  different strip widths. (a)~The minimum required current density to move the
  \BP{} past the notch at $x=300\,\mathrm{nm}$ increases with decreasing strip
  width. (b)~The \BP{} moves past the notch at a critical total current that is
  independent of the details of the geometry. In both subfigures, we can see
  that narrower strips compress the overall structure and allow the \BP{} to
  move closer to the notch or sample edge.}
\end{figure}

To better understand the effect of the notch size on the pinning, we
simulate strips with three different widths $w=100\,\mathrm{nm}$,
$w=125\,\mathrm{nm}$, and $w=150\,\mathrm{nm}$ for several different current
densities.
We keep the notch size of $n_y=30\,\mathrm{nm}$ used before.
Results are shown in Fig.~\ref{fig:single-notch-current-density}.

Figure~\ref{fig:single-notch-current-density}a shows the final position of the
\BP{} as a function of current density. Final positions below the notch tip at
$x=300\,\mathrm{nm}$, visualised by the grey dashed line, mean that the
\BP{} cannot move past the notch. Larger final positions mean that the \BP{} moves
past the notch. We can see that the minimum current density required to move the
\BP{} past the notch increases with decreasing strip width, as can be
expected because the fraction by which the overall structure
with the embedded \BP{} has to be compressed increases with decreasing strip width.

For large current densities, $J_x\geq 1.2\times10^{12}\,\mathrm{A/m}{}^2$, the \BP{} moves
past the notch for all strip widths. Here, we can see that the \BP{} can move
closer to the right sample boundary when the strip width decreases. This is
presumably a result of the fact that the large-scale magnetisation configuration
around the \BP{} is more compressed in narrower strips. It is in agreement with
our previous work~\cite{Lang2023}, where we find that the optimal distance
between \BP{}s in a configuration containing multiple \BP{}s also decreases with
decreasing strip width. It appears that the large-scale configuration around the
\BP{} wants to retain its approximately circular shape and reduces its radius
due to the narrowness (in $y$~direction) of the strip.

Figure~\ref{fig:single-notch-current-density}b shows the final position as a
function of total current through the $yz$~plane at the notch tip, the minimum of
the constriction. For similar geometries studied here, i.e.\ always a notch of
the same size at the same $x$~position, we find that the total current required
to move the \BP{} past the notch is approximately independent of the strip
width.

\subsection{Nanostrip with multiple notches}

Based on the previous results, we can develop a protocol to move one or multiple \BP{}s
past a series of notches in a controlled manner. The overall idea is as follows.
For weak current strengths, the \BP{} cannot move past a notch. Hence, we can
use a weak current to move a \BP{} to a defined position close to a notch.
Subsequently, we can apply a short, strong current pulse that pushes the \BP{}
past the notch. Afterwards, we can let the system relax (i.e.\ switch off the
current) or use a weak current to move the \BP{} to the next notch.

\begin{figure}
  \includegraphics[width=\linewidth]{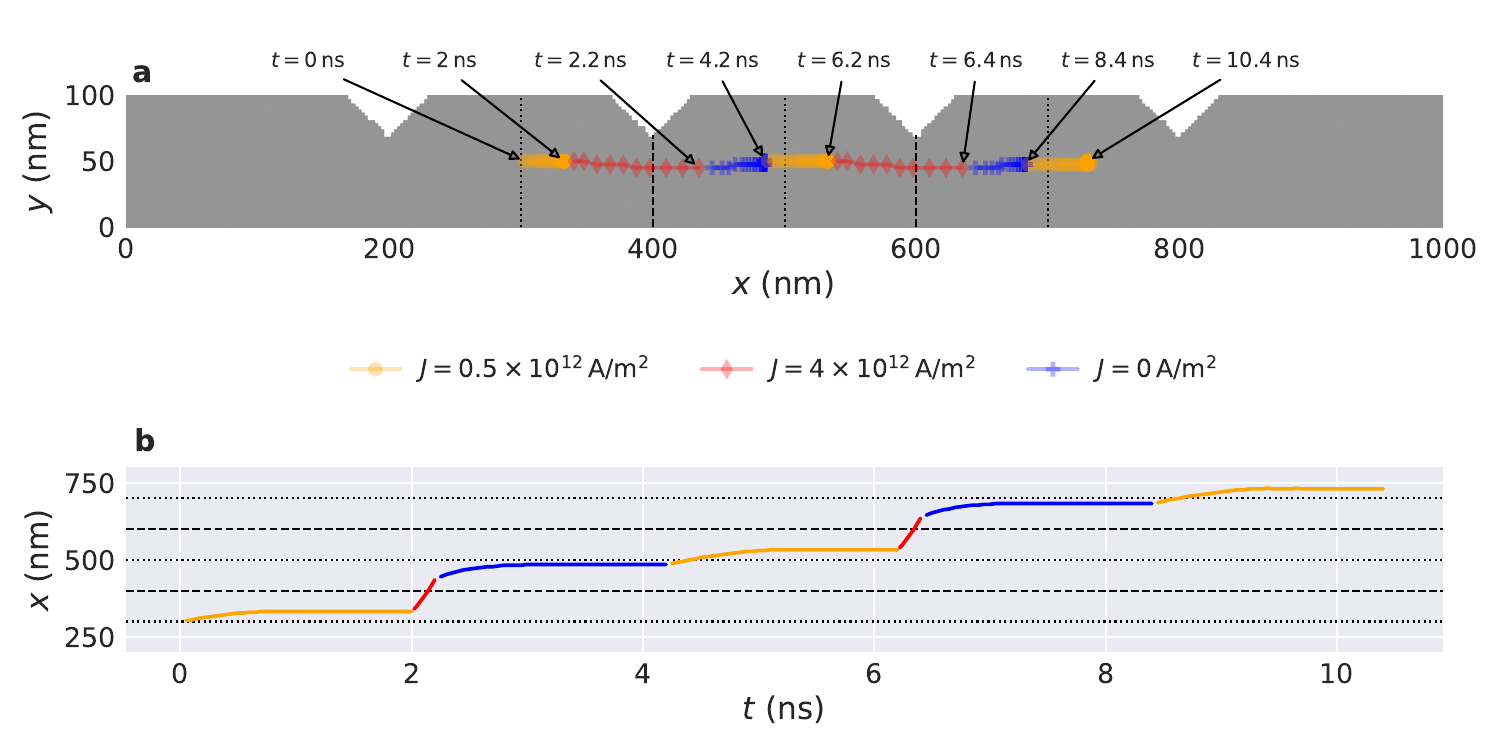}
  \caption{\label{fig:multi-notch-single-bp}Driving a single \BP{} past several
    notches using a series of current pulses of different strength. (a) Strip
    geometry and trajectory of the \BP{}. (b) Position of the
    \BP{} in $x$~direction as a function of simulation time. Dashed and dotted
    lines indicate the locations of the notches and centres of the wide areas in
  between, respectively.}
\end{figure}

Figure~\ref{fig:multi-notch-single-bp} demonstrates this process for a single
\BP{} in a strip with length $l=1000\,\mathrm{nm}$ and four notches at
$x=200\,\mathrm{nm}$, $x=400\,\mathrm{nm}$, $x=600\,\mathrm{nm}$, and
$x=800\,\mathrm{nm}$. We initialise the system in a configuration containing a
single \BP{} between the first two notches at $x=300\,\mathrm{nm}$.
Figure~\ref{fig:multi-notch-single-bp}a shows the strip geometry and the path of
the \BP{}, Fig.~\ref{fig:multi-notch-single-bp}b shows the $x$~position of the
\BP{} as a function of simulation time~$t$.

We use a three-step process to move the \BP{} past a notch. (i) We apply a
current $J_{x,\mathrm{(i)}}=0.5\times10^{12}\,\mathrm{A/m}{}^2$ to move the \BP{} to the next notch,
where it gets trapped. We use a pulse length $\Delta
t_{\mathrm{(i)}}=2\,\mathrm{ns}$ for this alignment of the \BP{} to the left of
the notch. From the time-resolved data we can see that the \BP{} stops moving
after $\Delta t\approx0.5\,\mathrm{ns}$. The longer pulse duration can be useful
to ensure that the \BP{} reaches the notch independent of its initial position.
Applying the pulse for ``too long'' does not affect the configuration due to the
trapping at the notch. (ii)~We apply a strong pulse with
$J_{x,\mathrm{(ii)}}=4\times10^{12}\,\mathrm{A/m}{}^2$ with pulse length $\Delta
t_{\mathrm{(ii)}}=0.2\,\mathrm{ns}$. This pulse pushes the \BP{} past the notch.
A short pulse duration is required to ensure that the \BP{} only moves past one
notch. (iii)~We remove the current and let the system relax for $\Delta
t_{\mathrm{(iii)}}=2\,\mathrm{ns}$. During this period, the \BP{} moves away
from the notch until it reaches its equilibrium position near the centre of the
region between the two notches, hereinafter called storage area. In the
simulation, the \BP{} reaches its equilibrium position after $\Delta t\approx
0.5\,\mathrm{ns}$ and does not move for the remainder of $\Delta
t_{\mathrm{(iii)}}$.

We can repeat steps (i) to (iii) to move the \BP{} past a series of notches, as
shown in Fig.~\ref{fig:multi-notch-single-bp}. Step (iii) is not strictly
required to achieve the desired motion past multiple notches. However, it
demonstrates several advantages. First, the \BP{} in the free system reaches an
equilibrium position in each storage area without an applied
current. Hence, successful operation only requires an applied current during a
short period of time. This reduces energy consumption of potential devices based
on this technology as external energy is only required to change the
configuration. Second, the position of the \BP{} at the end of step (ii) is not
critical. As long as the \BP{} has moved past the notch, the configuration will
automatically converge to a low-energy state with the \BP{} in the desired
storage area. This makes the approach more robust and less sensitive to, for
example, variations in current pulse duration and notch geometry.

\begin{figure}
  \includegraphics[width=\linewidth]{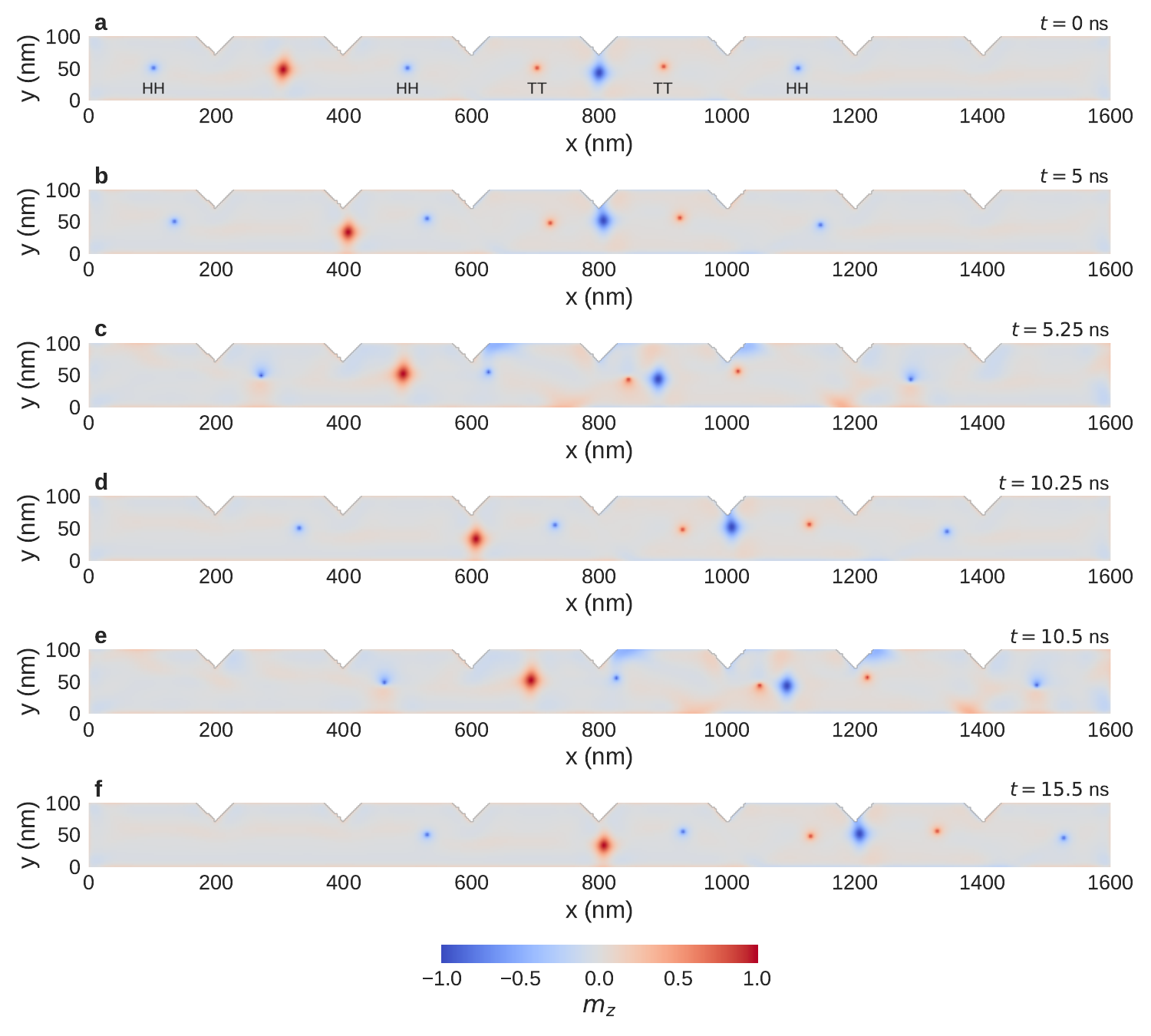}
  \caption{\label{fig:multi-notch-multi-bp}Movement of a configuration
    containing five \BP{}s past two notches by applying a series of current
    pulses of alternating lower and higher strength. The initial configuration
    and the distribution of the \BP{}s across storage areas is retained.}
\end{figure}

Figure~\ref{fig:multi-notch-multi-bp} demonstrates a similar process for a
configuration containing multiple \BP{}s in the configuration HH-HH-TT-TT-HH
(following the notation in~\cite{Lang2023}). We simulate a strip with length
$l=1600\,\mathrm{nm}$ containing seven notches.
Figure~\ref{fig:multi-notch-multi-bp}a shows the initial configuration with the
individual \BP{}s labelled.

The two large red and blue blobs at $x\approx300\,\mathrm{nm}$ and
$x\approx800\,\mathrm{nm}$ are antivortices that form between neighbouring
\BP{}s of the same type and which have significant magnetisation in the
$z$~direction. A more detailed study of the antivortices and their role in
a multi-\BP[-]{} system is provided in Ref.~\onlinecite{Lang2023}. For the following
discussion, it is sufficient to focus on the \BP{}s.

To simplify the simulation process, we only use steps (i) with $\Delta
t_{\mathrm{(i)}}=5\,\mathrm{ns}$ and (ii) with $\Delta
t_{\mathrm{(ii)}}=0.25\,\mathrm{ns}$. We start with (i) a weak current that
moves the \BP{}s towards the notches where they get trapped to the left of each
notch (Fig.~\ref{fig:multi-notch-multi-bp}b). The antivortices experience a
repelling force from the \BP{}s but do not significantly interact with the
notches. Subsequently, we apply (ii) the strong current pulse that pushes each
of the \BP{}s past the next notch to their right
(Fig.~\ref{fig:multi-notch-multi-bp}c). In the configuration at the end of this
pulse, at $t=5.25\,\mathrm{ns}$, we can see that individual \BP{}s have travelled
different distances during the pulse. After the next weak pulse
(Fig.~\ref{fig:multi-notch-multi-bp}d) each of the
\BP{}s is aligned to the left of the next notch, and all \BP{}s have the same
spacing to their notch. By applying the weak pulse for a sufficiently long
period, we ensure that all \BP{}s reach their aligned position near the next
notch before applying the next strong pulse. The cycle is repeated once more
(Fig.~\ref{fig:multi-notch-multi-bp}e and Fig.~\ref{fig:multi-notch-multi-bp}f),
and we can see, at $t=15.5\,\mathrm{ns}$, that each \BP{} has moved one notch
further (in comparison to Fig.~\ref{fig:multi-notch-multi-bp}d) in the direction of the applied current.

\subsection{T-shape geometry}

\begin{figure}
  \includegraphics[width=\linewidth]{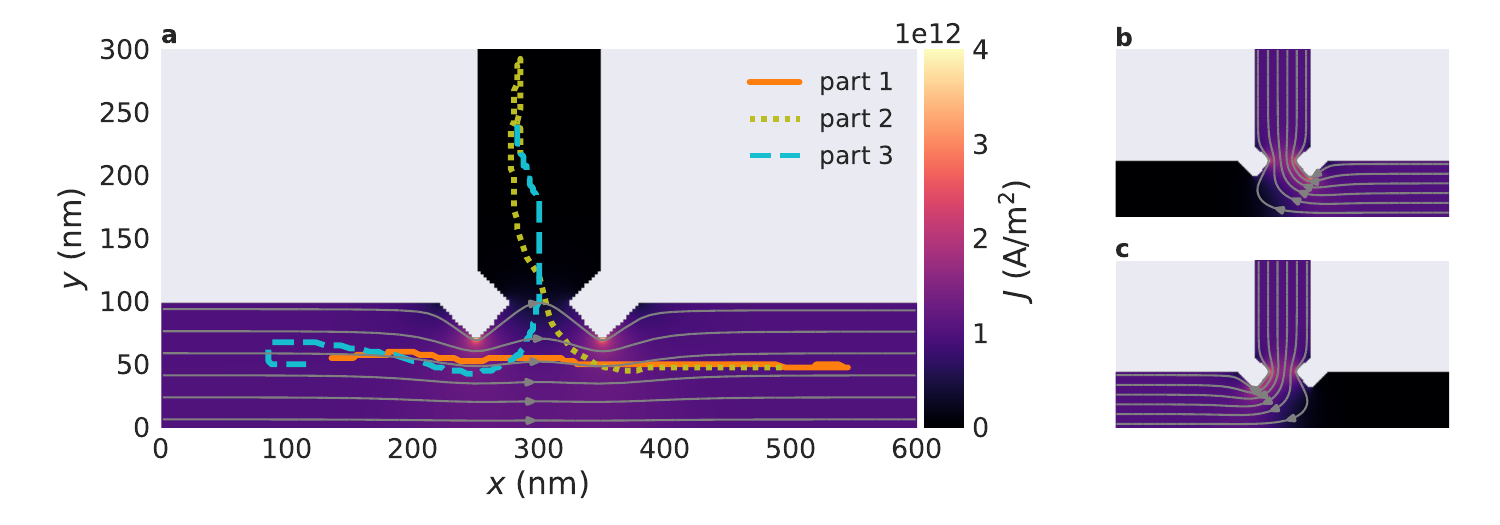}
  \caption{\label{fig:T-shape}Movement of a single \BP{} through a T-shaped
    structure when applying a series of current pulses along different
    directions. The \BP{} is initialised in the left storage area, its position
    during the simulation is shown with the thick line in~(a). The \BP{} first
    moves to the right storage area during part~1 with current applied from the left to
    the right as shown in the background of~(a), then to the top during part~2 with
    current applied from the right to the top as shown in~(b), and finally back
    to the left during part~3 with current applied from the top to the left as shown
    in~(c).}
\end{figure}

Finally, we study a single \BP{} in a T-shaped geometry with three
storage areas. The geometry is shown in Fig.~\ref{fig:T-shape}. It has the three
storage areas \emph{left} for $x<200\,\mathrm{nm}$, \emph{right} for
$x>400\,\mathrm{nm}$, and \emph{top} for $y>150\,\mathrm{nm}$. They are
separated by a total of four notches. We initialise the system in a state where
it contains a single \BP{} in the left storage area. We apply a series of
current pulses of varying strength between different pairs of strip ends to move
the \BP{} first to the right storage area, then to the top storage area, and
finally back to the left storage area.

Each part of the movement consists of three steps (i) long current pulse with
$J_{\mathrm{(i)}}=0.5\times10^{12}\,\mathrm{A/m}{}^2$ to move the \BP{} to the
notches where it gets stuck, (ii) short current pulse with
$J_{\mathrm{(ii)}}=5\times10^{12}\,\mathrm{A/m}{}^2$ to push the \BP{} past the
notches, and (iii) free relaxation. The weak pulse (i) is always applied for
$\Delta t_{\mathrm{(i)}}=2\,\mathrm{ns}$, during which the \BP{} moves towards the
notch and gets stuck well before the end of the simulation time. The duration of
the strong pulse depends on the part of the motion (see below). For the free
relaxation in step (iii) we simulate the time evolution of the system until it
reaches an equilibrium state (roughly for $\Delta t_{\mathrm{(iii)}}=2 -
4\,\mathrm{ns}$).

Figure~\ref{fig:T-shape}a shows the position of the \BP{} during the full cycle.
The three parts of the trajectory are shown with a solid orange line for the motion
left--right (part~1), a dotted olive line for the motion right--top (part~2), and a dashed
cyan line for the motion top--left (part~3). The background in Fig.~\ref{fig:T-shape}a
shows the magnitude of the current density for the current flowing from left to right,
streamlines show the current direction. Figures~\ref{fig:T-shape}b and~c show the
current flowing from the right to the top and from the top to the left,
respectively.

For the motion left--right we apply the strong pulse for
$\Delta t_{\mathrm{(ii)}}=0.7\,\mathrm{ns}$. We can see a slight overshoot at the end
of the pulse. During the free relaxation the \BP{} moves in $-x$~direction to
its equilibrium position at $x\approx500\,\mathrm{nm}$ near the centre of the
right storage area. For the motion right--top we apply the strong pulse for
$\Delta t_{\mathrm{(ii)}}=0.5\,\mathrm{ns}$. Here, we can see a strong overshoot with
the \BP{} nearly reaching the top sample boundary at $y=300\,\mathrm{nm}$.
During the free relaxation the \BP{} moves back to the central part of the
storage area at $y\approx200\,\mathrm{nm}$. For the final motion top--left we
apply the strong pulse for $\Delta t_{\mathrm{(iii)}}=0.7\,\mathrm{ns}$. After moving
past the two notches, the \BP{} gets deflected in $+y$~direction. This
deflection is a result of the uneven thickness of the two layers and the
deformed, off-centred \BP{} after it moves past the notches and the junction. We
have simulated a system with more similar layer thicknesses and find that the
effect decreases. During the free relaxation, the \BP{} moves back to the
central part of the storage area at $x\approx100\,\mathrm{nm}$. The slightly
different initial and final position in the left storage area result from the
differences in the initial and final magnetisation configuration due to the
series of applied current pulses and the nearly flat energy landscape for a
\BP{} inside a storage
area. Qualitatively similar behaviour can be seen in
Fig.~\ref{fig:multi-notch-single-bp}, where the free \BP{} also relaxes slightly
off-centred.

\section{Discussion}

Our micromagnetic simulations show that \BP{}s---which are equilibrium
configurations in two-layer FeGe nanostrips---can be moved with spin-polarised
currents. In uniform strips, \BP{}s move along the current direction without a Hall effect,
independent of their type. This is different from behaviour of
vortices~\cite{He2006, Shibata2006, Nakatani2008} or
skyrmions~\cite{Zang2011, Jiang2017a, Litzius2017}, and
also \BP{}s in other systems~\cite{Gong2021}. The straight motion is a special feature of the
two-layer system, in which each \BP{} is encapsulated between two vortices with
opposite polarisation. The vortices would be subject to deflection in opposite
direction but are strongly exchange-coupled across the layer interface. At the
\BP{} the forces are compensated, and thus the \BP{} moves in a straight line.
Different behaviour has been reported by Gong \emph{et al.}~\cite{Gong2021}, who
have studied current-induced dynamics of a chiral bobber~\cite{Rybakov2015,
  Zheng2018, Ahmed2018}, a skyrmion tube that ends in a \BP{}. They have
simulated thin FeGe films with a single material chirality with applied
spin-transfer
torque and observe a motion of the chiral bobber, and hence also the
\BP{}, with a current-dependent Hall effect.

In the system studied here, changes in geometry can be used to modify the
motion. We have demonstrated that below a threshold current density a \BP{} can
be trapped by a notch. The \BP{} holds its position because the current is
pushing it towards the notch constriction, but the \BP{} is repelled from the
boundary and does not want to deform,
and thus it cannot move past the notch. Using these competing effects, we can
choose appropriate current strengths to either neatly align a \BP{} at a notch
or move it past a notch. Using a series of current
pulses of different strength, we can move an array of \BP{}s of different type
through strips with multiple notches in a controlled fashion. They retain their
initial order and relative distance in terms of empty or occupied storage areas.
The additional antivortices between same-type \BP{}s do not impede this process,
despite not being trapped at the notches themselves. In each storage area we can distinguish
between three different local configurations: a HH \BP{}, a TT \BP{}, or no
\BP{}.

Furthermore, we have demonstrated that \BP{}s can be moved through more complex
geometries with multiple possible paths. We have demonstrated this for a
T-shaped geometry, where the \BP{} can move along either of the two possible
strip ends when being pushed towards the junction depending on the direction of
the applied current. In this system, we have added additional constrictions at
the junction to create three well-separated storage areas. The \BP{} in the free
system reaches an equilibrium position in each of the three storage areas. A
weak and subsequent strong pulse can be used to move the \BP{} between storage
areas.

In a larger system, these two building blocks---constrictions to restrict the
\BP[-]{} motion and junctions with multiple possible paths---can be combined.
Such a system can then host an array of \BP{}s and the geometrical constraints
can be use to manipulate the array, e.g.\ re-arrange \BP{}s with a series of
current pulses between different contacts. Successful operation will likely
require more device engineering.

Speculating about potential applications, a simple nanostrip containing a series
of \BP{}s can be used in a racetrack-like~\cite{Parkin2008} design, and the two
\BP[-]{} types can be used for binary data representation. The whole array of
\BP{}s could be moved with spin currents. Retaining equal spacing would not be
required because of the two different types that could encode ``0'' and ``1'',
which is not the case in skyrmion-based racetrack memories. In a more complex
setup we have shown that a series of notches in a strip can be used to create
confined ``storage areas''. \BP{}s can be distributed across these areas, where
each area can be in one of three states: occupied with a HH~\BP{}, occupied with
a TT~\BP{}, or not containing a \BP{}---with the potential to realise a ternary
storage device. The whole configuration could be moved through this strip using
a series of current pulses. Pinning of the \BP{}s at the notches helps
controlling the movement and ensures that the configuration is retained.

\section{Methods}
\subsection{Micromagnetic simulations}

We simulate rectangular two-layer nanostrips with opposite chirality (opposite
sign of $D$) in the two layers (Fig.~\ref{fig:strip-geometry}). We fix the thickness of the two layers to
$t_{\mathrm{b}} = 20\,\mathrm{nm}$ for the bottom layer and $t_{\mathrm{t}} =
10\,\mathrm{nm}$ for the top layer. We use a strip width of $w=100\,\mathrm{nm}$
unless indicated differently. We choose strip lengths that allow for enough
space for the \BP{}s to move and vary the strip length depending on the number
of \BP{}s. The energy equation:
\begin{equation}
  E = \int \mathrm{d}^3r \left( w_{\mathrm{ex}} + w_{\mathrm{dmi}} + w_{\mathrm{d}} \right)
\end{equation}
contains exchange energy density $w_{\mathrm{ex}}$, bulk Dzyaloshinskii-Moriya
energy density $w_{\mathrm{dmi}}$, and demagnetisation energy density
$w_{\mathrm{d}}$. The magnetisation dynamics is simulated using the
Landau-Lifshitz-Gilbert equation~\cite{Landau1935, Gilbert2004} with currents
modeled with the Zhang-Li model~\cite{Zhang2004}:
\begin{equation}
  \frac{\partial\mathbf{m}}{\partial t} = \gamma \mathbf{m} \times \mathbf{H}_{\mathrm{eff}}
  + \alpha \mathbf{m} \times \frac{\partial \mathbf{m}}{\partial t}
  -\mathbf{m} \times [ \mathbf{m} \times (\mathbf{u} \cdot \nabla)\mathbf{m} ]
  - \beta \mathbf{m} \times (\mathbf{u}\cdot\nabla)\mathbf{m},
\end{equation}
where $\gamma = 2.211 \times 10^5\,\mathrm{m/As}$ is the gyromagnetic
ratio, $\alpha$ is the Gilbert damping, and
\begin{equation}
\mathbf{u}=\frac{P\mu_{\mathrm{B}}}{eM_{\mathrm{S}}(1+\beta^2)}\ \mathbf{J} \label{eq:bp-motion:j-u}
\end{equation}
is the spin-drift velocity. Here, $\mathbf{J}$ is the electric
current density, $P$ is the polarisation, $\mu_{\mathrm{B}}$ the Bohr magneton,
$e$ the electron charge, and $\beta$ the non-adiabatic parameter. Material
parameters are based on FeGe~\cite{Beg2015}: $A =
8.87\,\mathrm{pJ}\,\mathrm{m}^{-1}$, $D = 1.58\,\mathrm{mJ}\,\mathrm{m}^{-2}$,
$M_{\mathrm{s}} = 384\,\mathrm{kA}\,\mathrm{m}^{-1}$, and
$\alpha=0.28$~\cite{Beg2017}. We use
spatially varying current densities $\mathbf{J}$ of different magnitudes in our simulations and
keep the other variables in Eq.~\ref{eq:bp-motion:j-u} fixed to $P=0.5$ and
$\beta=2\alpha = 0.56$.

All micromagnetic simulations are performed using Ubermag~\cite{Beg2022,
  Beg2017a} with OOMMF~\cite{Donahue1999} as the computational backend and
an extension for DMI for crystalclass T~\cite{Cortes-Ortuno2018,
  Cortes-Ortuno2018a}. We have generalised the Zhang-Li OOMMF extension, in
order to simulate current flowing in arbitrary directions. The modified
extension is available on GitHub~\cite{Lang2023a}.

As a starting point for all simulations we create rectangular nanostrips
containing \BP{}s at the desired positions, following a protocol that we
developed previously~\cite{Lang2023}. For rectangular nanostrips we can then
directly add the Zhang-Li term with a uniform current density to the dynamics
equation and simulate the \BP{} dynamics. For more complex geometries, we first
modify the nanostrip to have the desired shape and again minimise the energy to
start from a relaxed configuration. Then, we add the Zhang-Li current using the
current profile obtained form the finite-elements simulations outlined below
and simulate the \BP{} dynamics.

\subsection{Locating \BP{}s}

To analyse the \BP[-]{} motion we need to locate and track the individual
\BP{}s. We use a combination of two different methods to precisely locate the
\BP{}s. Tracking is then simply done based on the distance of \BP{}s in two
consecutive time steps. In all simulations the \BP{}s are clearly separated and
their number is kept fixed, hence identifying the individual \BP{}s using this
simple distance-based method is sufficient.

We first compute the approximate location of the \BP{}s (within cell accuracy)
based on a classification mechanism that we developed
previously~\cite{Lang2023}. We briefly summarise the method here. \BP{}s can be
identified as sources and sinks of the emergent magnetic field. To locate \BP{}s
along one direction we compute the flux of the emergent magnetic field through a
series of volumes that we increase along the respective direction. We find
quantised jumps in the total flux, whenever the volume includes an additional
\BP{}. Based on the sign of the jump we can determine the \BP[-]{} type. The
approximate location of the \BP{} is given by the upper integration boundary. To
locate a \BP{} in three dimensions we can repeat this calculation along
different directions, where we limit the integration volume along the directions
where we have already located the \BP{}. That way we can locate individual
\BP{}s in a configuration containing multiple \BP{}s.

To locate a \BP{} with sub cell size accuracy, we can compute the centre of mass
of the emergent magnetic field~$\mathbf{B}^{\mathrm{e}}$, defined as:
\begin{equation}
  \mathbf{r}_{\mathrm{BP}} = \frac{\int_V \mathrm{d}^3r\ \mathbf{r} \operatorname{div} \mathbf{B}^{\mathrm{e}}}{\int_V \mathrm{d}^3r\ \operatorname{div} \mathbf{B}^{\mathrm{e}}}
\end{equation}
where
\begin{equation}
  B^{\mathrm{e}}_i = \epsilon_{ijk} \mathbf{m} \cdot \left(\frac{\partial \mathbf{m}}{\partial r_j} \times \frac{\partial \mathbf{m}}{\partial r_k}\right).
\end{equation}
This method only works if the considered volume
contains a single \BP{}. Furthermore, magnetisation tilts at sample boundaries
can affect the result. To get reliable results, we first compute the approximate
location of all \BP{}s. Then, we can compute the precise location of each \BP{} by only
considering a small subvolume~$V$ around the approximate position. In the
majority of the work we only use the computationally less expensive locating method
with cell-size accuracy.

\subsection{Current profile}

We use Python libraries, which are part of the FEniCSx
project~\cite{Alnaes2014, Scroggs2022, Scroggs2022a}, to
numerically compute the current profile in non-rectangular nanostrips.
Additionally, we use Gmsh~\cite{Geuzaine2009} to create the irregular mesh
for the geometry.

For an electric conductor, according to the Ohm's law, the electric current density
$\mathbf{J}$ is defined via:
\begin{equation}
  \mathbf{J} = \sigma \mathbf{E} \\
\end{equation}
where $\mathbf{E}$ is the electric field and $\sigma$ the electric conductivity.
Further, the principle of charge conservation yields:
\begin{equation}
  \nabla \cdot \mathbf{J} = 0 \\
\end{equation}
According to Maxwell's equations, in the absence of a time varying magnetic
field, the electric field is conservative. Hence, it can be expressed in terms
of an electric potential as:
\begin{equation}
  \mathbf{E} = -\nabla V.
\end{equation}
Combining these equations, we obtain:
\begin{equation}
  \nabla \cdot (-\sigma\nabla V) = 0,
\end{equation}
Further, we assume as isotropic material, hence the conductivity is
a scalar. This gives a Laplace's equation:
\begin{equation}
  \nabla^{2}V = 0
\end{equation}
which we can solve numerically after defining suitable boundary conditions.

Figure~\ref{fig:strip-geometry} shows an example for a rectangular nanostrip
with a single notch. The calculated current density profile is shown on the top
surface. Figure~\ref{fig:single-notch-configurations}c shows a cut plane of
the same geometry. We assume a constant current flow through the left and right
sample boundary with strength $J_0$ in $+x$~direction and no current
flow through any of the other surfaces. Hence, the Neumann boundary conditions
can be expressed as:
\begin{equation}
  \frac{\partial V}{\partial n} =
  \begin{cases}
    -{J_0}/{\sigma}\qquad&\text{if } x=0\,\mathrm{nm}\text{ }\\
    \phantom{-}{J_0}/{\sigma}&\text{if } x=600\,\mathrm{nm}\text{ }\\
    \phantom{-}0&\text{else}
  \end{cases}.
\end{equation}
In the example we use $J_0=10^{12}\,\mathrm{A/m}{}^2$. The streamlines in
Fig.~\ref{fig:single-notch-configurations}c indicate the current direction,
colour the magnitude of $\mathbf{J}$. We obtain a uniform flux in $x$~direction
in the rectangular parts of the nanostrip. Near the notch, the current profile
changes as the current flows around the notch. This leads to a variation in the
current density with the maximum at the tip of the notch. To include the current
into the finite-difference micromagnetic simulations, we interpolate the
simulated current profile onto a cuboidal mesh. Visualisations in
Fig.~\ref{fig:single-notch-configurations} and Fig.~\ref{fig:T-shape} are shown
on the finite-difference grid used for the micromagnetic simulations.

\subsection{Computational science and data analysis}

For the preparation of simulation studies and data analysis we make use of the
ecosystem of Python-based open-source tools and libraries~\cite{
  Harris2020,  % numpy
  Hoyer2017,  % xarray
  Kluyver2016, Granger2021,  % Jupyter
  Hunter2007,  % matplotlib
  Rudiger2023,  % holoviews
  Sullivan2019}.  % pyvista

\section{Data availability}
All results obtained in this work can be reproduced from the repository in
Ref.~\onlinecite{Lang2023b} that contains software specifications and Jupyter
notebooks~\cite{Granger2021, Beg2021a} to
rerun all simulations and recreate all data and plots.

\bibliography{references}

\section{Acknowledgements}
This work was financially supported by the EPSRC Programme grant on Skyrmionics
(EP/N032128/1). The OpenDreamKit Horizon 2020 European Research Infrastructures
project (\#676541) has contributed to the Ubermag software, which was used heavily in
this work. We acknowledge the use of the HPC system at the Max Planck Institute
for the Structure and Dynamics of Matter, in the completion of this work.

\section{Author contributions}
M.L., M.B. and H.F. conceived the study.
M.L. performed the micromagnetic simulations and analysed the data.
S.P. performed the simulations of the current distribution.
M.L., M.B., S.H., S.P., and H.F. developed the Ubermag software package to drive the finite-difference solver OOMMF.
M.L. and H.F. prepared the manuscript.

\section{Competing interests}
The authors declare no competing interests.

\end{document}